\pdfoutput=1
\documentclass{aastex}
\shorttitle{APO and CARMA Observations of Mol 160}
\begin{document}

\title{Near-infrared and Millimeter-wavelength Observations of Mol 160: 
A Massive Young Protostellar Core}

\author{Grace Wolf-Chase\altaffilmark{1} and Michael Smutko\altaffilmark{2}}
\affil{Astronomy Department, Adler Planetarium, 1300 S. Lake Shore Drive,
Chicago, IL 60605}
\email{gwolfchase@adlerplanetarium.org}

\author{Reid Sherman and Doyal A. Harper}
\affil{Dept. of Astronomy \& Astrophysics, University of Chicago, 5640 S. Ellis Ave.,
Chicago, IL 60637}

\and

\author{Michael Medford}
\affil{Department of Physics \& Astronomy, Northwestern University, 2145 Sheridan Rd., Evanston, IL 60208}

\altaffiltext{1}{Dept. of Astronomy \& Astrophysics, University of Chicago, 5640 S. Ellis Ave.,
Chicago, IL 60637}
\altaffiltext{2}{Center for Interdisciplinary Exploration and Research in Astrophysics (CIERA)
\& Dept. of Physics \& Astronomy, Northwestern University, 
2145 Sheridan Rd., Evanston, IL 60208}

\begin{abstract}

We have discovered two compact sources of shocked H$_2$ 2.12-$\mu$m emission coincident with \object{Mol 160} (\object{IRAS 23385+6053}), a massive star-forming core thought to be a precursor to an ultracompact \ion{H}{2} region. The 2.12-$\mu$m sources lie within 2$^{\prime\prime}$ (0.05 pc) of a millimeter-wavelength continuum peak where the column density is $\ge$ 10$^{24}$ cm$^{-2}$. We estimate that the ratio of molecular hydrogen luminosity to bolometric luminosity is $>$ 0.2\%, indicating a high ratio of mechanical to radiant luminosity. CS J=2$\rightarrow$1 and HCO$^+$ J=1$\rightarrow$0 observations with CARMA indicate that the protostellar molecular core has a peculiar velocity of $\sim$ 2 km s$^{-1}$ with respect to its parent molecular cloud. We also observed 95 GHz CH$_3$OH J=8$\rightarrow$7 Class I maser emission from several locations within the core. Comparison with previous observations of 44-GHz CH$_3$OH maser emission shows the maser sources have a high mean ratio of 95-GHz to 44-GHz intensity. Our observations strengthen the case that \object{Mol 160} (\object{IRAS 23385+6053}) is a rapidly accreting massive protostellar system in a very early phase of its evolution.

\end{abstract}


\section{INTRODUCTION}

Studying the earliest evolutionary phases of massive protostars is difficult, because they are rare and their formation times are brief. Mol 160/IRAS 23385+6053 (hereafter Mol 160) is one of a number of sources identified by Molinari et al. (1996, 1998a, 2000, 2002) as possible precursors to ultracompact (UC)  \ion{H}{2} regions (Wood \& Churchwell 1989). They were selected on the basis of their cool IRAS colors, high infrared luminosities, absence of centimeter-wavelength continuum emission, H$_2$O maser activity, ammonia line emission, millimeter continuum observations, and presence of energetic molecular outflows. Based on follow-up observations with the Owens Valley Radio Observatory (OVRO) and the Infrared Satellite Observatory (ISO), Molinari et al. (1998b, hereafter M98) suggested that Mol 160 might be the first bona fide example of a massive Class 0 protostar, an object still accreting matter at a high rate from a massive envelope. 

It is generally accepted that outflows from low-mass young stellar objects (YSOs) are driven by magnetic stresses in accretion disks that eject some of the inflowing matter and carry off angular momentum (Pudritz \& Norman 1983, 1986; Lovelace et al. 1987; K\"onigl 1989; Pelletier \& Pudritz 1992; Wardle \& K\"onigl 1993; Safier 1993; Paatz \& Camenzind 1996; Shu et al. 2000). Optical and near-infrared observations often reveal jets and Herbig-Haro objects that show in detail the interaction of the underlying winds with the ambient medium.

If massive stars are built up through similar accretion processes, one might expect them to have outflows that are scaled-up versions of their low-mass counterparts. However, observationally linking jets and outflows with their sources is challenging, even in regions forming intermediate-mass stars (e.g., Wolf-Chase et al. 2003). The problems are compounded for massive stars, which often form in tight clusters, lie in heavily obscured regions at distances greater than several kiloparsecs, and may be evolving on dynamical time scales less than a few thousand years.

Observations of molecular hydrogen lines in the near infrared are particularly valuable, since they directly measure kinetic energy deposited by high-speed shocks and can be observed with large ground-based telescopes at angular resolutions similar to those that can be achieved with millimeter wavelength interferometry. In this paper, we present narrowband, 0.7$^{\prime\prime}$-resolution, near-infrared images of Mol 160 made with the Apache Point Observatory (APO) 3.5-m telescope and 2$^{\prime\prime}$-resolution 3-mm continuum and line observations with the Combined Array for Research in Millimeter-wave Astronomy (CARMA). The near-infrared measurements were made as part of an emission-line survey for shocked molecular hydrogen gas associated with massive protostars (Wolf-Chase et al. 2011, in preparation). In the Mol 160 region, we detected a number of compact molecular hydrogen emission-line sources. Surprisingly, two of the brightest lie within 2$^{\prime\prime}$ of a millimeter-wavelength continuum peak where the hydrogen column density is $\ge$ 10$^{24}$ cm$^{-2}$. Our CARMA measurements include CS, a tracer of dense gas, HCO$^+$, a tracer of cloud kinematics, and a Class I maser transition of CH$_3$OH, a tracer of shock-excited gas. Our observations support the hypothesis that Mol 160 is a massive protostellar system in a very early phase of evolution, but call into question some of the specific conclusions drawn from earlier outflow studies of this object (see \S 4.3.1).

Prior work on Mol 160 includes observations covering a wide range of wavelengths and angular resolutions, from the arcminute scales of the IRAS measurements to the arcsecond scales of interferometric observations. This can potentially lead to ambiguities in the association of source names with specific physical systems. In this paper, we follow the nomenclature generally used in the literature (e.g., Wang et al. 2011 and references therein). We use ``clumps'' to refer to parsec-scale molecular cloud structures and``cores'' to the smaller structures ($\sim$ 0.1 pc) that may further collapse and coalesce into stars or groups of stars. We will use the term ``Mol 160'' to narrowly denote the protostar (or protostellar cluster) and the dense molecular core from which it is forming (diameter $\le$ 6$^{\prime\prime}$) and ``Mol 160 region'' to refer to the arcminute-scale area that also encompasses potentially related structures including larger molecular clouds and two \ion{H}{2} regions.

\section
{OBSERVATIONS AND DATA REDUCTION}

\subsection{Apache Point Observatory: NICFPS}

We obtained narrowband, near-infrared images using the Near-Infrared Camera and Fabry-Perot Spectrometer (NICFPS) on the Astrophysical Research Consortium (ARC) 3.5-m telescope at APO in Sunspot, NM. NICFPS uses a 1024 $\times$ 1024 Rockwell Hawaii HgCdTe detector with 18-$\micron$ pixels. On the APO 3.5 m, this yields a pixel scale of 0.273$\arcsec$ pixel$^{-1}$ and gives a field of view of 4.6$\arcmin \times 4.6\arcmin$ (Vincent et al. 2003). We used the following filters: K$_s$ (500 s), H$_2$ 2.12-$\micron$ (1800 s), H$_2$ 2.25-$\micron$ (1200 s), and a narrow-band H$_2$-continuum filter centered on 2.13 $\micron$ (1200 s). Because of weather and scheduling constraints, the H$_2$ 2.12-$\micron$ and K$_s$ data were recorded on 2005 Jun 17 and the remaining data were obtained on 2007 Nov 21. Seeing conditions were good the first night ($\sim0.7\arcsec$ at 2.2 $\micron$) and average the second night ($ \sim1.3\arcsec$ at 2.2 $\micron$).

All images were processed using IRAF and standard reduction techniques (e.g. Joyce 1992). We dithered the telescope between each exposure in a five-point pattern, and used the median of the resulting images to subtract the sky background from each image. To produce final images in each filter, we used stars to register the reduced images and then took the median of the aligned frames.

We produced a continuum-subtracted H$_2$ 2.25-$\micron$ image by taking the difference between H$_2$ 2.25-$\micron$ and H$_2$-continuum images. Ideally, we would have done the same for the H$_2$ 2.12-$\micron$ data, but it was not possible to obtain H$_2$-continuum filter data during our 2005 run. 
Additionally, the instrument electronics were upgraded between the two runs, which resulted in different read noise and gain levels on the chip that complicated the comparison of 2005 and 2007 data.
For these reasons, we chose to use K$_s$ data taken on the same night and thus with the same electronics and seeing conditions as the H$_2$ 2.12-$\micron$ data, scaling the K$_s$ image so that the stars disappeared when we subtracted the K$_s$ image from the H$_2$ 2.12-$\micron$ image. Both continuum subtraction techniques worked well on all but the brightest stars, which show small residual artifacts.

We performed aperture photometry on compact sources using the DAOPHOT package in IRAF. To avoid source confusion while still maintaining good signal-to-noise, we chose an aperture radius of 8 pixels (2.2$\arcsec$), which is roughly 3 times the FWHM of the stars in the H$_2$ 2.12-$\micron$ images.

The data were calibrated by comparing the instrumental magnitudes of several dozen stars to their K$_s$ magnitudes in the 2MASS point source catalog. For each filter, we calculated a zero point correction between our instrumental magnitudes and the 2MASS K$_s$ magnitudes, then computed fluxes using:

$$ F_{Jy} = F_0 \times 10^{-Kmag/2.5}  \eqno{(1)}  $$
$$ F_{line} = F_{Jy} \times 10^{-26} \times d\nu \eqno{(2)} $$

\noindent where F$_0 =$ 666.7 Jy and $d\nu = c  d\lambda/\lambda^2$.  For the H$_2$ 2.12-$\micron$ filter, $d\lambda = 6.93$ nm and for the H$_2$ 2.25-$\micron$ filter, $d\lambda = 7.6$ nm (Apache Point Observatory 2010).

WCS coordinates were calibrated using 2MASS point sources. The 2MASS positions are accurate to within 0.1$^{\prime\prime}$ for K$_s$ $<$ 14 (Skrutskie et al. 2006). Given the seeing conditions ($\sim$ 0.7$^{\prime\prime}$) during acquisition of the NICFPS data  and pixel size of 0.27$^{\prime\prime}$, we estimate that our positions are accurate to within $\pm$ 0.3$^{\prime\prime}$. 

\subsection{CARMA}

We observed a field around the Mol 160 core approximately 1.5$^{\prime}$ across with CARMA on 2010 Mar 23 and 24 during three tracks. The interferometer was in the C configuration with baselines ranging from 26-370 meters. We simultaneously observed the continuum and 3 spectral lines. At the time, CARMA was configured to observe in both upper and lower sidebands of three spectral bands, and a fourth was being integrated into the system.

For the first track, two bands were configured for maximum continuum bandwidth, covering 468 MHz each, for a total of 1.872 GHz (including both sidebands). A third observed CH$_3$OH J=8$\rightarrow$7 at 95.169 GHz in the upper side band with a 31-MHz bandwidth and 63 channels, resulting in a velocity resolution of 1.538 km s$^{-1}$. The fourth band observed HCO$^+$ J=1$\rightarrow$0 at 89.189 GHz in the lower side band with 62-MHz bandwidth and 383 channels, giving a resolution of 0.547 km s$^{-1}$. There were calibration problems with this band; the upper side band data were corrupted, so the fourth band was not used during the other tracks.

For the second and third tracks, we used only three bands. One observed continuum, for a total of 936-MHz bandwidth. The band observing CH$_3$OH was unchanged. The third band observed the HCO$^+$ line in the lower side band and CS J=2$\rightarrow$1 in the upper side band at 97.981 GHz. It was set to a 62-MHz bandwidth with 63 channels, yielding 3.283 km s$^{-1}$ resolution for HCO$^+$ and 2.988 km s$^{-1}$ resolution for CS.

The passband calibrator was 3C84 for the first track, 1927$+$730 on the second track, and 3C454.3 on the third. The flux calibrator was Uranus for the first and second track and MWC349 on the third. The phase calibrator was 0102$+$584 for all tracks.

In our analysis, we combined CS and HCO$^+$ data from the second and third tracks and CH$_3$OH and continuum data from all three tracks. The total integration time on source was 2.35 hours for the second and third tracks and 4.75 hours for all three tracks. The size (FWHM) of the synthesized beam was 1.982$^{\prime\prime}\times$1.811$^{\prime\prime}$. We reduced and calibrated the data using the MIRIAD data reduction package (Sault et al. 1995). We used an intermediate Brigg's visibility factor to balance sensitivity to compact structure in the Mol 160 core with suppression of sidelobes that might otherwise distort clump-scale emission. Coverage of the u-v plane over the multiple tracks contains numerous shorter baselines, so we expect good qualitative rendering of structures $\le$30$^{\prime\prime}$ in size, though some fraction of the flux will be missing.

\section
{RESULTS}

\subsection{H$_2$ Emission in the Mol 160 Region}

Our H$_2$ 2.12-$\micron$ and continuum-subtracted H$_2$ 2.12-$\micron$ images are shown in Figures 1 \& 2, respectively. Positions and designations of the 24 $\micron$ point sources (crosses) presented in Table 1 of Molinari et al. (2008a, hereafter M08) are plotted in Figure 2. Source ``A'' lies close to the Mol 160 molecular core. Some of the pure line emission in Figure 2 arises from fluorescently excited gas in photo-dissociation regions (PDRs) associated with two \ion{H}{2} regions (indicated by the large ovals in Figure 1) and some from shock-excited gas.  In a catalog hosted by the Joint Astronomy Centre in Hawaii, Davis et al. (2010) have established a numbering scheme for ``molecular hydrogen emission-line objects'' or ``MHOs'', compact emission-line sources thought to be associated with outflows. We have identified 10 candidate MHOs and numbered them according to this scheme. Their positions are plotted with open diamonds in Figure 2 and identified by the last two digits of the MHO numbers.

Table 1 lists the MHO designations (column 1), their positions (columns 2 \& 3), and their 2.12-$\micron$ line fluxes (column 4). None of the MHOs were detected in the H$_2$ 2.25-$\micron$ filter. Using equations (1) \& (2) to calibrate the faintest stars in the H$_2$ 2.25-$\micron$ image against stars also present in the 2MASS catalog, we derive a 3$\sigma$ upper limit for H$_2$ 2.25-$\micron$ line intensity of $9.49 \times 10^{-19}$ W m$^{-2}$. Column 5 lists the ratios of the 2.12-$\micron$ fluxes to the 2.25-$\micron$ upper limit. The difference in extinction at 2.12 $\micron$ and 2.25 $\micron$ amounts to only a tenth of a magnitude (assuming A$_{\lambda} \sim {\lambda}^{-1.7}$), with a corresponding brightness ratio of $\sim$1.1, which would change the numbers in column 5 only slightly. The expected 2.12-$\micron$/2.25-$\micron$ line ratio for UV excitation (as in PDRs) is $\sim$ 1.9, and $\sim$ 7.7 for shocks (Black \& van Dishoeck 1987; Gredel \& Dalgarno 1995). Shock excitation is consistent with the observed ratios for all the MHOs and is very likely for the brightest.

Star formation is clearly occurring throughout the Mol 160 region.
MHOs 2921 \& 2922 are of particular interest because of their proximity to the Mol 160 3.2-mm continuum peak.  At the present time, there is no evidence linking any of the other MHOs to the Mol 160 core itself. MHO 2922 is extremely compact, appearing almost stellar in the 2.12-$\micron$ image (see Figure 3a). Comparison with stellar images of comparable peak intensity yields a deconvolved source size of 0.7$^{\prime\prime} \times 0.6^{\prime\prime}$. Assuming a distance of 4.9 kpc to Mol 160 (M98)\footnote{Zhang et al. (2005) used a distance of 6.9 kpc, and M08 suggest the distance may be closer to 8 kpc.}, this corresponds to a physical size of $\sim$ 3430 AU. MHO 2921 is slightly larger and more irregular in appearance, particularly at low intensity levels. It is displaced to the south of the peak position of the 3.2-mm continuum source, and there is a very dark region symmetrically displaced to the north, with a low-intensity arc of H$_2$ 2.12-$\micron$ emission at its northern edge, as can be seen in the continuum-subtracted 2.12-$\micron$ image (Figure 3b). Fitting a two-dimensional Gaussian to the bright central region of MHO 2921 yields a deconvolved size of  1.4$^{\prime\prime} \times 0.7^{\prime\prime}$.

Subtracting the Gaussian fits to MHO 2921 and MHO 2922 from the continuum-subtracted H$_2$ image of Figure 3b results in the image shown in Figure 3c. The residual emission has the form of a ring of approximately the same diameter as the outer contours of the  -53.7 km s$^{-1}$ CS and -52.5 km s$^{-1}$ HCO$^+$ peaks seen in Figure 4. The ring also coincides with a number of CH$_3$OH
maser sources.
The only feature of comparable intensity within the ring is a patch of emission near the positions of Masers 3 and 4 (see \S 3.2.2). In Figure 3, the relative intensities at the peaks along the ring and the peaks at MHO 2921 \& MHO 2922
are roughly 1:4:10. Note that the intensities of the dark area and arc north of the main continuum source are the same in Figures 3b and 3c. That is, they are not artifacts of the subtraction.

\subsection{CARMA Results}

\subsubsection{HCO$^+$, CS, and 3.2-mm Continuum Emission}

The 3.2-mm (93.6 GHz)
continuum emission peaks at the position 23$^h$40$^m$54.5$^s$, +61$^{\circ}$10$^{\prime}$28.1$^{\prime\prime}$. The beam size was 2.0$^{\prime\prime}$ $\times$ 1.8$^{\prime\prime}$, with a position angle of 40$^{\circ}$. A two-dimensional Gaussian fit to our continuum data yields a size of 3.3$^{\prime\prime}$ $\times$ 2.6$^{\prime\prime}$, with a position angle of 40$^{\circ}$. The deconvolved source size is 2.7$^{\prime\prime}$ $\times$ 1.9$^{\prime\prime}$ with a position angle of 35$^{\circ}$. We measured a peak intensity of 10.8 mJy beam$^{-1}$. Integrating under the Gaussian fit yields a total flux of 21.6 mJy. Our position lies within 0.4$^{\prime\prime}$ of the positions measured by M98 and Fontani et al. (2004, hereafter F04) at 3.4mm (88 GHz) and 3 mm ($\sim$100 GHz), respectively. M98 measured a peak intensity of 9 mJy beam$^{-1}$ in a 4.1$^{\prime\prime}$ $\times$ 3.5$^{\prime\prime}$ beam, a total flux of 19 mJy, and derived a deconvolved source size of 4.5$^{\prime\prime}$ $\times$ 3.6$^{\prime\prime}$. F04 measured a peak flux of 8 mJy beam$^{-1}$ in a 2.3$^{\prime\prime}$ $\times$ 1.9$^{\prime\prime}$ beam, a total flux of 12.4 mJy (integrated to the 3$\sigma$ contour), and a deconvolved source size of 1.5$^{\prime\prime}$. Our measured flux at 3.2 mm is consistent with the SED for Mol 160 presented in Fig. 4 of M08, and does not change their conclusions regarding envelope properties or integrated bolometric luminosity (3170 L$_{\odot}$, assuming a distance of 4.9 kpc).

Maps of the three CS and three HCO$^+$ channels containing significant signal are shown in Figure 4 and Figure 5. The bandwidths of each channel are $\sim$ 3 km s$^{-1}$ (3.283 km s$^{-1}$ for HCO$^+$ and 2.988 km s$^{-1}$ for CS). The central frequencies of the HCO$^+$ channels lie near the boundaries of the CS channels. Figure 4 shows the data at full angular resolution. In Figure 5, the data have been smoothed using a Gaussian kernel with a radius of 4$^{\prime\prime}$, to emphasize structure with scale lengths between those of the full-resolution interferometer data and measurements made with single-dish telescopes. Contours of velocity-integrated CS emission are plotted on the H$_2$ image in Figure 1. 

Although the interferometer data do not recover all the flux at larger scales, the data have been smoothed on scales to which CARMA is still sensitive, so the structures seen in the smoothed maps are either real emission, or real emission contaminated by side-lobe artifacts.  To test for the latter, we compared the smoothed, velocity-integrated CS emission with the CARMA PSF smoothed to the same radius. We found no correlation of CS emission features with side-lobe features, and therefore conclude that the emission features seen in the smoothed maps are real. Qualitatively, Figure 5 is more sensitive to parsec-scale ``clumps'' and Figure 4 to sub-parsec-scale ``cores''. Furthermore, we note that the CS clumps in the smoothed maps correlate well with depressions in the extended H$_2$ emission (Figure 1), as would be expected for dense clumps in the foreground of a bright PDR background.
% Changes to above paragraph address the artifact issue raised by the referee.

In the smoothed maps, clump-scale emission is most evident in the CS channel centered at -50.7 km s$^{-1}$ and the HCO$^+$ channels centered at -49.2 and 52.5 km s$^{-1}$. The most prominent clump is the one containing Mol 160. Near the position of the 3.2-mm continuum peak, the velocity distribution is broader and ``bluer'' than at other positions in the clump, and the compact nature of the Mol 160 core is most evident in the CS channel centered at -53.7 km s$^{-1}$ and the HCO$^+$ channel centered at -52.5 km s$^{-1}$.

\subsubsection{CH$_3$OH Masers}

Kurtz, Hofner, \& \'Alvarez (2004, hereafter KHA) detected 44-GHz CH$_3$OH Class I maser emission at four positions near the Mol 160 core with a synthesized beam of $2.01^{\prime\prime} \times 1.25^{\prime\prime}$ with the VLA in its D configuration. We found 95.169-GHz CH$_3$OH J=8$\rightarrow$7 emission at all the positions they reported and also at a position south of the Mol 160 core continuum peak. We obtained a copy of the 44-GHz  data (S. Kurtz, private communication), and compared it with our CARMA observations. Examination of the individual channel maps revealed 44-GHz emission at the position of our southern 95-GHz source and at slightly lower but statistically significant intensities at several other positions. The channel maps also show that the KHA Source 3 position lies between two partially resolved sources, one to the north and one to the east of the Mol 160 core continuum peak. At the 0.166 km s$^{-1}$ resolution of the 44-GHz data, many of the sources shift in position from channel to channel, particularly Source 2, which extends in velocity across 6 channels and traces a path several arcseconds long. In Table 2 we list the peak positions and velocities of all the 44-GHz sources. We have extended the nomenclature used by KHA to include the three additional masers with intensities below the cut of their survey data and to reflect the resolution of their Source 3 into two spatial components.

To more directly compare the intensities of the sources at 95 GHz and 44 GHz, we velocity-binned the KHA data to the channel resolution of the 95-GHz data and smoothed the higher angular resolution VLA data to match the CARMA beam. The resulting maps are shown in Figure 6. Table 2 compares the properties of the 95-GHz and 44-GHz sources. The positions listed in Table 2 correspond to positions of peak single-channel intensity of the 44-GHz observations, which display more structure than the 95-GHz observations because of their higher angular resolution. The large diamonds in the panels 
identify the four KHA positions. Small diamonds indicate the positions of the sources with intensities lying below the KHA cutoff (Sources 5, 6, and 7) and the two partially-resolved components of Source 3 (3N and 3E). The crosses in the panels show the positions of peak single-channel 95-GHz intensity. Column 7 in Table 2 lists the peak single-channel 44-GHz intensities. Columns 8 and 9 comprise, respectively, the 44-GHz intensities integrated over a velocity range corresponding to the 95-GHz channel bandwidth and the 95-GHz peak intensities at the positions of Column 1 and 2. Column 10 lists the ratios of the 95-GHz intensities to the smoothed, velocity-integrated 44-GHz intensities.

\section{DISCUSSION}

In their review of high mass star formation, Zinnecker and Yorke (2007) suggest four evolutionary phases. (1) Compression: the formation of dense, cold, gravitationally bound cores with masses $\ge$ 100 M$_{\odot}$ from a larger molecular cloud by gravo-turbulent cloud fragmentation. (2) Collapse: gravitational collapse of portions of the core into one or more quasi-hydrostatic protostellar embryos, accretion disks, and infall envelopes. (3) Accretion: accretion of matter onto protostellar objects, eventually leading to one or more massive stars. (4) Disruption: disruption of the natal core by stellar winds, outflows, ionizing radiation, or supernovae.

In practice, linking details of these proposed evolutionary categories to observations of massive star-forming regions is complicated by many factors.
The total radiant luminosity of a massive protostellar core may arise from a combination of Kelvin-Helmholtz emission from quasi-static protostellar embryos, protostellar and disk accretion shocks, and shocks associated with outflows or relative motions of core components. During the earliest evolutionary phases, there could also be significant contributions from external heating sources. The relative importance of each could vary dramatically and perhaps erratically with time, depending on the detailed dynamics of what may be a complex and chaotically evolving system. The expected conversion efficiency of gas to stars is sufficiently small that the mass of a core may not change dramatically until it enters the disruption phase (see, e.g., the theoretical tracks of core mass vs. luminosity in Molinari et al. 2008b). However, its temperature, density structure, chemistry, and kinematics may undergo significant alterations throughout the compression, assembly, and accretion phases. The ratio of mechanical luminosity from outflows to total radiant luminosity may also vary in a complicated way.
 In a general sense, high accretion rates probably correlate with high outflow rates, but kinetic energy can be stored and released asynchronously. Additionally, outflows in massive star-forming regions may be produced through different mechanisms. 
 
Two distinctly different types of outflows have been observed in some of the nearest massive star-forming regions.
Whereas the outflow from IRAS 20126$+$4104 appears to be a scaled-up version of the disk-mediated accretion
outflow scenario associated with low-mass YSOs (Caratti o Garatti et al. 2008), 
outflow from the Orion BN/KL complex has been attributed to the explosive disintegration of a massive star cluster
about 500 years ago (e.g., Rosenthal et al. 2000; Zapata et al. 2009; Bally et al. 2011).
Since most massive star-forming regions (including Mol 160) are too distant to fully resolve outflow structure with current instruments, the frequency of occurrence of these fundamentally different types of outflow events, and their effects on 
evolving protoclusters, remains to be seen. For example, Mol 160 is approximately ten times farther away than OMC-1. At this distance, the entire BN/KL complex, including both the ``18 km s$^{-1}$'' SiO outflow and the higher velocity gas associated with the shocked H$_2$ ``bullets'' or ``fingers''  (Allen and Burton 1993; Rosenthal et al. 2000; Colgan et al. 2007), would span $\le$6$^{\prime\prime}$, similar to the size of the Mol 160 core.

 Theoretical models can offer some insights but are poorly constrained by actual observations. 
The observational problem at hand is how detailed studies of Mol 160 and other young objects can illuminate the phenomenology of massive star formation and place constraints on possible models. 
 
\subsection{Environment of Mol 160}

The molecular cloud containing Mol 160 was studied by F04. They observed C$^{18}$O J=1$\rightarrow$0, C$^{17}$O J=1$\rightarrow$0, and C$^{17}$O J=2$\rightarrow$1 with the IRAM 30-m telescope with 23$^{\prime\prime}$, 22$^{\prime\prime}$, and 11$^{\prime\prime}$ beams, respectively. At the position of the 3-mm continuum peak, they found there were two distinct kinematic components, centered at -50.5 and -47.8 km s$^{-1}$. Maps over a 40$^{\prime\prime} \times 40^{\prime\prime}$ area showed that the -47.8 km s$^{-1}$ emission peaked south of the position of the Mol 160 continuum peak and that the peak of the -50.5 km s$^{-1}$ emission was extended in a N-S direction and displaced about 4$^{\prime\prime}$ to the west. They suggested that the lack of a prominent feature at the position of the Mol 160 core might be explained by depletion of molecular gas on grains in the cool, high-density core. They also mapped NH$_3$ (1,1) over the same region with the VLA, using a 4.3$^{\prime\prime} \times 3.7^{\prime\prime}$ beam. They found the emission was extended in a N-S ridge parallel to the CO peak, but displaced only 1$^{\prime\prime}$-2$^{\prime\prime}$ to the west of Mol 160. At the declination of Mol 160, there was actually a depression in the ridge, with the peak intensity occurring $\sim$ 5$^{\prime\prime}$ to the south, again suggesting depletion in the core. In NH$_3$, they saw only the -50.5 km s$^{-1}$ component, and they concluded that the interferometer was resolving out the -47.8 km s$^{-1}$ component seen in single-dish observations with the Effelsberg 100-m telescope. Their CO maps also showed that the -47.8 km s$^{-1}$ emission was more extended E-W than the -50.5 km s$^{-1}$ component. Henceforth, we will refer to these components as the -50.5 and -47.8 km s$^{-1}$ clouds.

On a larger scale, Molinari et al. (2002) observed 3.6-cm continuum emission from two extended \ion{H}{2} regions lying to the west and east of Mol 160. They named them VLA 1 and VLA 2, respectively. Their ionization fronts almost intersect at a position just to the east of the Mol 160 core (see the large ovals on Figure 1). 
They appear to be bounded along portions of their ionization fronts by dense molecular clouds. In such cases, the high optical depths and densities toward the cloud result in high-surface-brightness H$_2$, polycyclic aromatic hydrocarbon (PAH), and dust emission from layers parallel to the ionization front, leading to particularly intense emission along tangent lines of sight. For the purposes of this discussion, we will broaden the conventional definition of the term PDR to include not only the zone just outside the ionization front but also the high-density zone inside the front where grains flowing into the \ion{H}{2} region from the molecular cloud are exposed to high radiation densities and emit strongly at mid-infrared wavelengths.

Contours of velocity-integrated CS emission are plotted on the H$_2$ image in Figure 1.
We smoothed the CS data using a Gaussian kernel with a 4$^{\prime\prime}$ radius in order to highlight emission from parsec-scale structures generally referred to as ``clumps'' (Wang et al. 2011 and references therein). Most of the flux seen in the integrated map is also present in the single velocity channel centered at -50.7 km s$^{-1}$. In the H$_2$ images, one can see dark nebulosity overlying a brighter background of fluorescent emission from PDRs excited by the stars ionizing the \ion{H}{2} regions. 
A similar morphology can be seen in the 15-$\micron$ ISO image of M98. The molecular clump containing the Mol 160 core lies in a particularly dark band whose E-W extent is comparable to the width of the -50.5 km s$^{-1}$ CO cloud. It seems likely that the -50.5 km s$^{-1}$ cloud and the Mol 160 core are associated with this dark nebulosity and that they lie on the near side of the \ion{H}{2} regions and their associated PDRs. The CS clumps lie along the eastern side of the dark band, to the east of the CO ridge and just to the west of the intense arc of PDR emission at the western boundary of VLA 2.

M08 observed Mol 160 with MIPS on Spitzer at 24 $\micron$ and 70 $\micron$. The prominent arcs of emission in their 24-$\micron$ map all have analogs in our 2.12-$\micron$ image. The extended 24-$\micron$ and fluorescent H$_2$ radiation probably arise from contiguous regions in the PDR, within the first few optical depths at FUV wavelengths. Our arcsecond-resolution H$_2$ data suggest that much of the flux that they attribute to embedded 24-$\micron$ point sources (the positions for which are shown in Figure 2) could arise from knots of intense PDR emission. Separating truly point-like mid-infrared sources will require significantly better angular resolution than possible with Spitzer. The exciting stars of VLA 1 and VLA 2 may as yet be undetected.

It is likely that much of the 70-$\micron$ emission also arises from the PDRs. The overall morphologies of the Spitzer 24-$\micron$ and 70-$\micron$ maps made by M08 are similar. One can plausibly ascribe most of the differences to the fact that the hotter grains that dominate the 24-$\micron$ emission lie close to or within the \ion{H}{2} regions, where the stellar radiation is most intense. The principal exception is the far infrared peak near the position of Mol 160. Even in that case, the relative importance of internal and external heating is uncertain. The 70-$\micron$ flux peaks $\sim$ 5$^{\prime\prime}$ northeast of the Mol 160 continuum source. The discrepancy could result from positional uncertainties, as they suggest, but it is perhaps as likely that the shift is real and a significant contribution to the peak intensity in the 22$^{\prime\prime}$ Spitzer beam comes from nearby PDR emission, as is the case in arcminute-scale observations of the Trapezium/BNKL region in Orion (Harper 1974). Because the column density of the  Mol 160 core is sufficiently high to absorb mid-infrared flux from the PDR, there could also be a significant external contribution to the heating of the core. For these reasons, the value M08 derive for the radiant luminosity of the Mol 160 core should be considered an upper limit.

\subsection{The Mol 160 Core}

F04 have summarized estimates of the mass, column density, volume density, and temperature of the Mol 160 core derived from both line and continuum data taken with single dish telescopes and interferometers at angular resolutions between 1.5$^{\prime\prime}$ and 18$^{\prime\prime}$. For a region with diameter of $\sim$4$^{\prime\prime}$-8$^{\prime\prime}$ and an assumed distance of 4.9 kpc, they find  a kinetic temperature of 26 K, masses of $\sim$100-400 M$_{\odot}$, H$_2$ column densities of $\sim$ 1.5-4 $\times$10$^{24}$ cm$^{-2}$, and space densities of $\sim$ 3.4-16 $\times$10$^{6}$ cm$^{-3}$, with higher numbers from continuum data than from molecular lines. They note that there is evidence that molecules are depleted onto grains in the densest part of the core and suggest that the higher values should be preferred. For a region 1.3$^{\prime\prime}$-1.9$^{\prime\prime}$ in diameter ($\sim$ 0.03 pc), they derive a kinetic temperature of 42 K, masses of 15-150 M$_{\odot}$, H$_2$ column densities of $\sim$3.3-15.0 $\times$ 10$^{24}$ cm$^{-2}$, and space densities of $\sim$ 1.7-6.0 $\times$10$^{7}$ cm$^{-3}$, again with the highest numbers derived from continuum data. From the source SED, they estimate a temperature of 40 K. M08 derived a similar temperature of 37 K by using the automatic SED fitting tool provided by Robitaille et al. (2007), and a bolometric luminosity of 3170 L$_{\odot}$.  They showed that these values place Mol 160 firmly into the protostar category on the evolutionary diagram developed by Molinari et al. (2008b). We note that their estimate of the bolometric luminosity of Mol 160 should be considered an upper limit, due to emission from the nearby,
overlapping PDRs.
Taken together, these observations and models indicate Mol 160 is in an evolutionary stage that is a precursor to a hot core. 

In spite of the uncertainties, it seems reasonable to conclude that the mass of the core is sufficient to form one or more massive stars. Furthermore, the high column densities suggest that massive stars {\it should} form once the core begins to collapse. Recent theoretical work predicts a mass column density threshold $\ge$ 1 g cm$^{-2}$ for massive star formation (Krumholz \& McKee 2008; Krumholz et al. 2010). Regions where intermediate mass stars are thought to be forming typically have mass column densities $\sim$ 0.1 - 0.5 g cm$^{-2}$ (e.g., Arvidsson et al. 2010; Wolf-Chase et al. 2003; Wolf-Chase, Walker, \& Lada 1995). In the Arvidsson et al. (2010) study, estimates were made from peak column densities derived from single-antenna millimeter-wave observations of regions at different distances, implying a `clump' rather than `core' scale for many of these regions; nevertheless, the derived values are significantly smaller than those obtained for UC \ion{H}{2} regions at comparable distances. In the Wolf-Chase et al. (1995, 2003) studies, the 2264 S1 core in the Mon OB1 dark cloud has a mass column density of  $\sim$ 0.27 g cm$^{-2}$ on $\sim$ 0.1 pc scale. Spitzer observations indicate 2264 S1 contains at least 10 low-mass protostars (Young et al. 2006; Teixeira et al. 2006). Taking 4.0 $\times$ 10$^{24}$ cm$^{-2}$ as a lower limit for the column density of the compact Mol 160 core (see Tables 5 \& 6 in F04), the mass column density is  $\ge$ 16 g cm$^{-2}$. 

M98 noted that the Mol 160 core was slightly elongated to the southeast and to the north at 3.4 mm. 
Using a $0.94^{\prime\prime} \times 0.76^{\prime\prime}$ beam at a wavelength of 1.3 mm, F04 resolved the millimeter continuum source into two components. The principal component was centered at the position previously determined by M98. The flux density per beam and integrated flux density of the smaller source were, respectively, 50\% and 8\% of those of the primary.  Our 3.2-mm continuum observations do not resolve these two components, but are slightly extended in the direction of the secondary peak (see contours on Figure 3c). Also, the position of the secondary source coincides with the position of MHO 2922. If the mass within a 1.5$^{\prime\prime}$ diameter region centered on the primary continuum peak is $\sim$ 150 M$_{\odot}$, as F04 suggest, and if the mass scales with the 1.3-mm continuum flux, the mass associated with the secondary peak would be $\sim$ 12 M$_{\odot}$.

The Mol 160 region was not covered in the Spitzer GLIMPSE survey (Benjamin et al. 2003; Churchwell et al. 2009), 
but has been observed by WISE. Figure 7 is a 3-color WISE image (3.4, 4.6, \& 12-$\micron$ bands) of the Mol 160 region. As in the color scheme chosen for the GLIMPSE survey, objects that are unusually bright in the 4.6-$\mu$m band appear to be green. In the GLIMPSE data, ``extended green objects'' or ``EGOs'' have been linked to shocked H$_2$ emission from massive outflow candidates (Cyganowski et al. 2008). MHOs 2921 \& 2922 are the only MHOs listed in Table 1 that are detected as EGOs in the WISE image. Furthermore,
Mol 160 is {\it only} detected in the WISE 4.6-$\mu$m band and was not seen in the continuum by ISO at 15 $\mu$m (M98). Hence it is likely, as for the EGOs, that the 4.6-$\mu$m emission is dominated by spectral lines from shocked molecular gas.
 
\subsection{Outflow from Mol 160}

Observations of millimeter-wave, H$_2$, and CH$_3$OH maser emission independently trace outflowing gas from Mol 160. Taken together, these observations point to the extreme youth of Mol 160 and suggest new avenues to explore
in establishing an evolutionary sequence for massive star formation.

\subsubsection{Millimeter-wave Spectroscopy}
    
M98 and Molinari et al. (2002) presented evidence for energetic molecular outflows in Mol 160. The wings of their SiO spectrum at the position of Mol 160 form a broad plateau in velocity space. The wings have a relatively large amplitude compared to the intensity in the central frequency channel, and the intensity drops sharply at the edges of the plateau. The HCO$^+$ wings are much smaller, compared to the central peak. The blue wing tapers down smoothly from the systemic peak within $\sim$ ~10 km s$^{-1}$. The red wing is broader, with a dip at -46 km s$^{-1}$ and a peak at -43 km s$^{-1}$. The blue and red HCO$^+$ outflow maps of M98 peak $\sim$ 1$^{\prime\prime}$ to the east of the continuum peak, and the blue peak (integrated from -63 km s$^{-1}$ to -53 km s$^{-1}$) is much stronger and broader than the barely detected red peak (integrated from -47 km s$^{-1}$ to -37 km s$^{-1}$). The frequency range over which they integrated to make their ``blue'' outflow map overlaps a third of our -52.5 HCO$^+$ channel and nearly all of our 53.7 km s$^{-1}$ CS channel, so it is likely that their map is dominated not by outflow gas but by the blue wing of the systemic core gas. On the other hand, the morphologies of their maps of the SiO wings are similar to each other. The peak of the blue lobe lies $\sim$ 0.5$^{\prime\prime}$ south of the continuum peak, and the red peak lies $\sim$ 1$^{\prime\prime}$ to the north. Based on the small angular separation between the lobes, M98 argued that the outflow is essentially parallel to the line of sight, ruling out the possibility that their failure to detect 15-$\mu$m emission with ISO was caused by geometrical effects such as an edge-on dust disk and thus implying that the flow is completely contained within the optically thick core.

In our HCO$^+$ data, we do see a NE-SW velocity gradient across the core (cf. Figure 4 and the color image in Figure 8). The channel maps of the 1.5 km s$^{-1}$ data from Track 1 show the same trend, but the instrumental problems with the correlator prevent a more precise estimate of the velocity dispersion. One can set a rough upper limit of $\sim$6 km s$^{-1}$ from the span between the central velocities of our -49.2 and -55.8 km s$^{-1}$ HCO$^+$ channels. This is much less than the $ >$20 km s$^{-1}$ extent of the SiO plateau and could plausibly result from systematic motions established during the compression/collapse phase of core evolution rather than protostellar outflows. We suggest that kinematics of the HCO$^+$ and CS gas can best be understood in terms of a model in which the compact, high-surface-brightness molecular core associated with the Mol 160 continuum source is blueshifted by  $\sim$2 km s$^{-1}$ from a parent clump with systemic velocity of -50.5 km s$^{-1}$. Most of the flux from the compact core lies within the -52.5 km s$^{-1}$ HCO$^+$ channel and -53.7 km s$^{-1}$ CS channel. The source brightness distributions in these two channels are similar and their outer contours are roughly congruent with the low-intensity 2.12-$\micron$ halo around MHO 2921 and MHO 2922 seen in Figure 3. The observed blueshift of the core gas with respect to the -50.5 km s$^{-1}$ cloud would be consistent with compression by a background \ion{H}{2} region.

Our observations are consistent with the assertion by M98 that the extent of the SiO outflow is small compared to the size of the core but suggest an alternative possibility for the geometry. 
The symmetrical displacements of MHO 2921 to the south of the continuum peak position and an opposing dark area to the north is reminiscent of the simulated near-infrared images of wide-angle outflows constructed by Zhang \& Tan (2011). See, for example, the example with an outflow inclination of 60$^{\circ}$ in their Figure 9. Their models assumed the source of the near-infrared radiation from the arc was continuum radiation from the protostar but would also be consistent with an intense source of line emission located close to the center of the core. Such a scenario would imply that H$_2$ emission associated with the red SiO outflow lobe is hidden by dust in the waist of the accretion envelope (and/or a disk or pseudodisk).

From their SiO observations, M98 deduced an outflow mass of $\sim$ 20 M$_{\odot}$, a dynamical timescale of $\le 6 \times 10^3$ yr, a mass loss rate of 3.5 $\times 10^{-3}$ M$_{\odot}$ yr$^{-1}$ and a mechanical luminosity of 22 L$_{\odot}$. They based these estimates on the assumptions that the outflow angle was 30$^{\circ}$ with respect to the line of sight and that the extent of the outflow was equal to the radius of the molecular core. In light of the small angular displacement of MHO 2921 from the continuum peak, the mechanical luminosity and mass loss rate could be larger and the time scale shorter, indicating an even younger, more powerful outflow.

\subsubsection{H$_2$ Luminosity}

Until spectra of MHOs 2921 \& 2922 are available, we can only estimate the effects of extinction. Assuming that the 2.12-$\micron$ flux is $\sim$ 5\% of the total ro-vibrational H$_2$ emission (e.g., Caratti o Garatti 2006, 2008), the combined H$_2$ luminosity for the two MHOs is $3.4 \times 10^{-3} \times$ (D/4.9 kpc)$^2 \times 10^{0.4 A_{2.12}}$ L$_{\odot}$, where D is the distance in kpc and $A_{2.12}$ is the extinction at 2.12 $\micron$. Because of extinction, the actual luminosities of the 2.12-$\micron$ sources are probably much greater than the observed luminosities. The average column densities, N(H$_2$),  of the large -50.5 km s$^{-1}$ cloud inferred from large-beam measurements with single-dish telescopes range from 1$\times 10^{23}$ to 3.6$\times 10^{23}$ cm$^{-2}$. The values derived from interferometric measurements at higher angular resolution are larger by an order of magnitude or more. Conservatively assuming a column density of N(H$_2$) = 5$\times 10^{22}$ cm$^{-2}$, an amount equal to half the lowest value derived from large-beam measurements, leads to a 2.12-$\micron$ extinction of 5.1 and a combined H$_2$ luminosity of 7.4 L$_{\odot}$ for MHO 2921 and MHO 2922. This is 0.23\% of the bolometric luminosity inferred from the SED of the continuum source (L$_{bol} = 3170$ L$_{\odot}$, M08) and is comparable to the mechanical luminosity inferred by M98 from the high-velocity SiO outflow. 

Caratti o Garatti (2006) found a relationship between total H$_2$ luminosity and L$_{bol}$  for low-mass YSOs.
Caratti o Garatti (2008) suggested that this relationship could tentatively be extended to high-mass YSOs based on their
results for IRAS 20126$+$4104. They further suggested that this relationship might be taken as evidence 
that outflows from massive YSOs are similar in nature to those from lower-mass YSOs
(i.e., produced by disk-mediated accretion). However, the Orion BN/KL complex, which has a bolometric luminosity of 10$^5$ L$_{\odot}$, and a total H$_2$ luminosity of 120 L$_{\odot}$ (Rosenthal et al. 2000), would also fit their relationship. In this kind of impulsive event there is no reason to expect strict proportionality between the bolometric luminosity of the source and either the kinetic energy of the outflow or its rate of dissipation in shocks. Therefore, caution must be exercised
in attempting to draw conclusions about the nature of outflows from massive YSOs from the L$_{H_2}$/L$_{bol}$ relationship alone. We note that the L$_{H_2}$/L$_{bol}$ ratio for Mol 160 conservatively places Mol 160 a factor of two above the L$_{H_2}$/L$_{bol}$ relationship determined by Caratti o Garatti (2006, 2008) and the discrepancy could be larger if (a) the internal heating is smaller; (b)  the extinction has been underestimated; or (c) the ratio of total H$_2$ to H$_{2.12}$ luminosity has been underestimated (e.g., if there is significant emission from cooler shocked hydrogen, cf. Caratti o Garatti 2006, 2008).

As noted in \S 4.3.1, MHO 2922 is coincident with a secondary continuum peak (F04). It is possible that MHO 2922 emission arises from a distinct outflow from the central source, a secondary accretion center, or an object ejected from the core. Present observations do not allow us to distinguish between the possibilities; however, we note that this does not change our conclusion that Mol 160 has a very high L$_{H_2}$/L$_{bol}$ ratio, since this ratio was computed based on the estimated bolometric luminosity for the entire core, and our computed ratio is a conservative lower limit.

\subsubsection{CH$_3$OH Masers}

Class I CH$_3$OH masers are frequently and perhaps exclusively found near young sources. They are well-correlated with molecular outflows in massive star forming regions and are thought to be collisionally pumped (Cyganowski et al. 2009 and references therein; Fontani et al. 2010). Schnee \& Carpenter (2009) found a strong correlation between the presence of compact 3-mm continuum emission and 95-GHz Class I CH$_3$OH maser emission. In contrast, they detected no 3-mm continuum emission toward UC \ion{H}{2} regions lacking maser emission, suggesting that the masers are signposts of an early stage in the evolution of a massive protostar before an expanding UC \ion{H}{2} region has destroyed the accretion disk.

DeBuizer et al. (2009) note that the relatively low velocities of CH$_3$OH masers, combined with their occasional locations slightly offset from the outflow axis, suggest they arise in ``systemic'' gas in outflow cavity walls. The velocities of the Mol 160 masers vary from -49.0 to 52.2 km s$^{-1}$, and the most intense emission comes from sources nearest the continuum peaks and at velocities near $\sim$ 52 km s$^{-1}$. The maser positions seem to lie along two principal axes (see Figure 8), one approximately north-south through the main continuum source and one roughly along a line between the main and secondary continuum sources. It is tempting, but probably premature, to associate them with the walls of a wide-angle outflow cavity. The source could be significantly more complex than a single central source and a single bipolar outflow. The maser sources are as yet unresolved, spectrally or spatially, but they are bright enough to permit observations at higher spectral and spatial resolution that could significantly improve our understanding of the structure, kinematics, and excitation of the outflows.

In their survey of CH$_3$OH masers, Fontani et al. (2010) found that
sources for which both 44-GHz and 95-GHz Class I masers were observed, 
had similar spectra, confirming a common physical origin. They also concluded that the 95-GHz line is intrinsically fainter, based on their detection rates. The mean value of the ratio of 95-GHz to 44-GHz line intensity in Table 2 is 4.4. Some of the individual sources studied by Fontani et al. (2010) also have intensity ratios this large, suggesting that their detection statistics must result not from the intrinsic line formation process but from the probability of finding sources with conditions favoring the 95-GHz transition. Perhaps the ratio will prove to be a useful diagnostic for the shortest-lived (e.g., very early) phases of massive star formation.

 \subsection{Mol 160: Current Status \& Future Explorations}

Taken together, the core and outflow properties of Mol 160 strongly suggest that this object is in an early phase of the 3rd stage (accretion) outlined for massive star formation at the beginning of this section (Zinnecker \& Yorke 2007). The temperature, mass, mass column density, and luminosity of Mol 160 are consistent with physical properties expected for a precursor to a hot core, before winds and radiation from embedded protostars have removed a significant amount of the accreting envelope, and while the protostar mass and luminosity are still increasing. The presence of a very compact outflow, which is independently confirmed via millimeter-wave spectroscopy, H$_2$ shocks, and CH$_3$OH masers, also points to the extreme youth of one or more embedded, accreting, massive objects.

Because we expect the youngest massive protostars to be rare, more refined selection methods are highly desirable. A high value of L$_{H_2}$/L$_{bol}$ may be a useful discriminant, but the example of Mol 160 also highlights some of the difficulties. Near- and mid-infrared spectroscopy are needed for more accurate estimates of extinction and L$_{H_2}$. It is also likely that massive accreting objects will be located in or near regions containing other products of massive star formation such as bright PDRs. For these sources, bolometric luminosities can easily be overestimated. High-angular-resolution, multicolor, far-infrared observations with Herschel or the Stratospheric Observatory for Infrared Astronomy (SOFIA) will help reduce confusion in complex regions. Systematic searches in infrared dark clouds more clearly separated from PDRs (e.g., for compact Herschel sources and GLIMPSE ``green objects'' more pointlike than EGOs) may also be helpful. The Milky Way Project is producing a new catalog of PDRs in the galactic plane using citizen scientist identifications from Spitzer GLIMPSE/MIPSGAL images (Simpson et al. 2011) that should be useful in this regard. High-angular-resolution observations of CH$_3$OH Class I maser line ratios may also prove to be a useful diagnostic for physical conditions characteristic of very young sources, one that will become even more valuable with the deployment of advanced receivers on CARMA and as the Atacama Large Millimeter/submillimeter Array  (ALMA) comes on line.

\section
{SUMMARY AND CONCLUSIONS}

\begin{enumerate}

\item 
We have discovered two bright, compact, molecular hydrogen emission-line objects in a dense molecular core suspected to contain a massive protostellar object undergoing rapid accretion. In a 2.12-$\micron$ continuum-subtracted image taken in 0.7$^{\prime\prime}$ seeing, MHO 2921 has a deconvolved size of 1.4$^{\prime\prime} \times 0.7^{\prime\prime}$ and lies $\sim$ 0.5$^{\prime\prime}$ south of the main 3.2-mm continuum source (a projected distance of $\sim 0.012 \times (D/4.9 kpc)$ pc). MHO 2922 is almost starlike in appearance, with a deconvolved size of 0.7$^{\prime\prime} \times 0.6^{\prime\prime}$. It lies $\sim$ 2$^{\prime\prime}$ east of the main continuum peak and within $\sim$ 0.5$^{\prime\prime}$ of a previously reported secondary continuum peak. The observed line fluxes of MHOs 2921 \& 2922 are each $\sim  2.4\times 10^{-18}$ W m$^{-2} \times (D/4.9 kpc)^2$.

\item
The star-forming molecular core appears to be kinematically distinct from its surroundings. It is blueshifted by $\sim$ 2 km s$^{-1}$ from a clump of dense gas with multiple cores and a systemic velocity of -50.5 km s$^{-1}$. The clump is in turn part of a larger -50.5 km s$^{-1}$ molecular cloud (east-west FWHM $\sim$ 19$^{\prime\prime}$) previously mapped in C$^{18}$O and C$^{17}$O with the single-dish 30-m IRAM telescope. In our 2.12-$\micron$ image, the large cloud can be seen as a dark band silhouetted against bright extended emission from photo-dissociation regions associated with two background \ion{H}{2} regions. The clump lies at the eastern edge of the large cloud, $\sim$ 4$^{\prime\prime}$ to the west of the N-S ridge of peak C$^{18}$0 emission.

\item
We detected CH$_3$OH J=8$\rightarrow$7 95 GHz Class I maser emission peaks
at three positions. In two locations, one north and the other east of the main continuum source, the emission peaks near previously observed 44-GHz maser emission. Closer inspection of the 44-GHz data also revealed emission at the position of the southern 95-GHz maser and several additional positions. The mean ratio of 95-GHz/44-GHz intensities at positions with statistically significant emission at both frequencies is 4.4.

\item
In addition to the compact MHOs, there is a halo of low-intensity diffuse 2.12-$\mu$m emission contiguous with the Mol 160 core. Its outer diameter is $\sim$ 5.4$^{\prime\prime}$, approximately the same as the outer contours of the molecular core and the extent of the CH$_3$OH maser emission. The present data cannot discriminate between direct emission from shocks associated with the masers and scattered emission from a central source. However, as in the case of the MHOs, the diffuse emission seems to be pure line emission, a conclusion supported by the fact that the source was only detected by WISE in its 4.6-$\micron$ band. There is no evidence for either direct or scattered continuum emission from embedded protostars. A dark depression in the halo emission north of the continuum peak position suggests a geometry in which H$_2$ line radiation from the red lobe of a bipolar outflow is hidden by dust in the waist of a tilted accretion envelope.

\item
Because of extinction, the actual luminosities of the 2.12-$\micron$ sources are probably much greater than the observed luminosities. What we believe to be a conservative lower limit to the extinction leads to an estimate of 7.4 L$_{\odot}$ 
for the combined H$_2$ luminosity of MHO 2921 and MHO 2922. This is 0.23\% of the bolometric luminosity inferred from the SED of the continuum source and about twice the value predicted from an extrapolation of the L$_{H_2}$/L$_{bol}$ relation derived for lower-luminosity MHOs. It is also of the same magnitude as the mechanical luminosity inferred from the high-velocity SiO outflow and about 6\% the H$_2$ luminosity of the explosive outflow in the BN/KL region of Orion.

\item
These results strengthen but do not yet prove the hypothesis that Mol 160 is a precursor to an ultracompact \ion{H}{2} region still in a very early evolutionary phase dominated by accretion. The current radiant luminosity of the source is less than expected for a B0.5 star, but the high mass surface density of the source is consistent with conditions favoring the formation of stars with equal or higher mass. The existing observations strongly suggest the presence of bright unresolved structure within the source. Additional infrared and millimeter wavelength observations at higher spatial and spectral resolution can be expected to yield much stronger constraints on the kinematics, luminosity, and structure of the source.

 \end{enumerate}

\acknowledgments

This research is based on observations obtained with the Apache Point Observatory 3.5-meter telescope, which is owned and operated by the Astrophysical Research Consortium. We particularly thank the 3.5-meter Observing Specialists for their assistance in acquiring the data. Support for CARMA construction was derived from the Gordon and Betty Moore Foundation, the Kenneth T. and Eileen L. Norris Foundation, the James S. McDonnell Foundation, the Associates of the California Institute of Technology, the University of Chicago, the states of California, Illinois, and Maryland, and the National Science Foundation. Ongoing CARMA development and operations are supported by the National Science Foundation under a cooperative agreement, and by the CARMA partner universities. Special thanks go to Stan Kurtz, for supplying us with his VLA CH$_3$OH 44 GHz maser data, and Scott Schnee and Claudia Cyganowski for helpful conversations regarding CH$_3$OH masers. We also thank the anonymous referee for many helpful suggestions to improve this paper. This research has made use of SAOImage DS9, developed by the Smithsonian Astrophysical Observatory. We acknowledge use of the NASA/IPAC Infrared Science Archive, which is operated by the Jet Propulsion Laboratory, California Institute of Technology, under contract with the National Aeronautics and Space Administration. This publication makes use of  data products from the Two Micron All Sky Survey, which is a joint project of the University of Massachusetts and the Infrared Processing and Analysis Center/California Institute of Technology, funded by the National Aeronautics and Space Administration and the National Science Foundation; and the Wide-field Infrared Survey Explorer, which is a joint project of the University of California, Los Angeles, and the Jet Propulsion Laboratory/California Institute of Technology, funded by the National Aeronautics and Space Administration. GW-C and MM were funded in part through NASA's Illinois Space Grant Consortium, and the authors gratefully acknowledge support from the Brinson Foundation grant in aid of astrophysics research at the Adler Planetarium.

\clearpage
\begin{figure}
\plotone{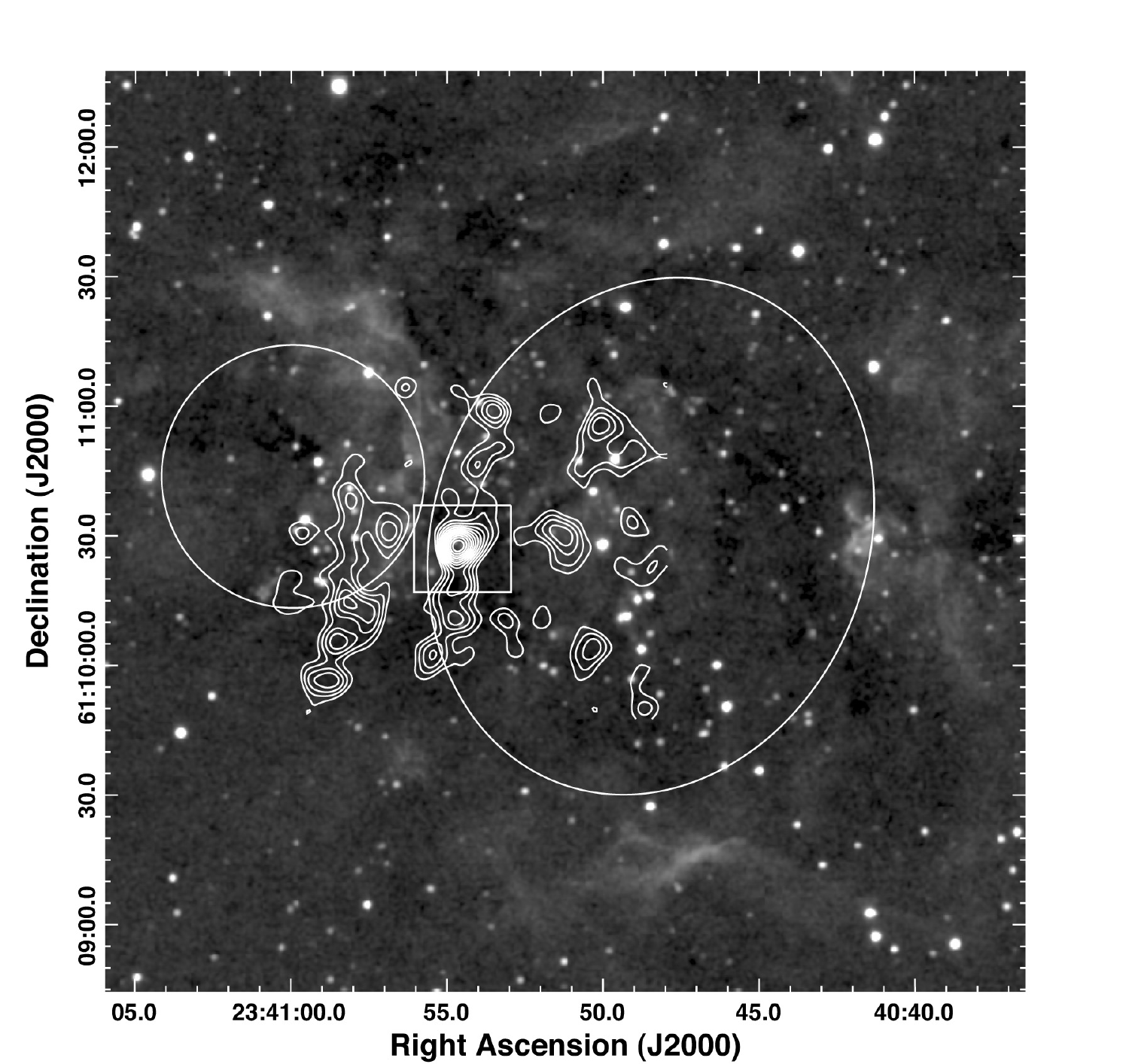}
\end{figure}

\clearpage
\begin{figure}
\plotone{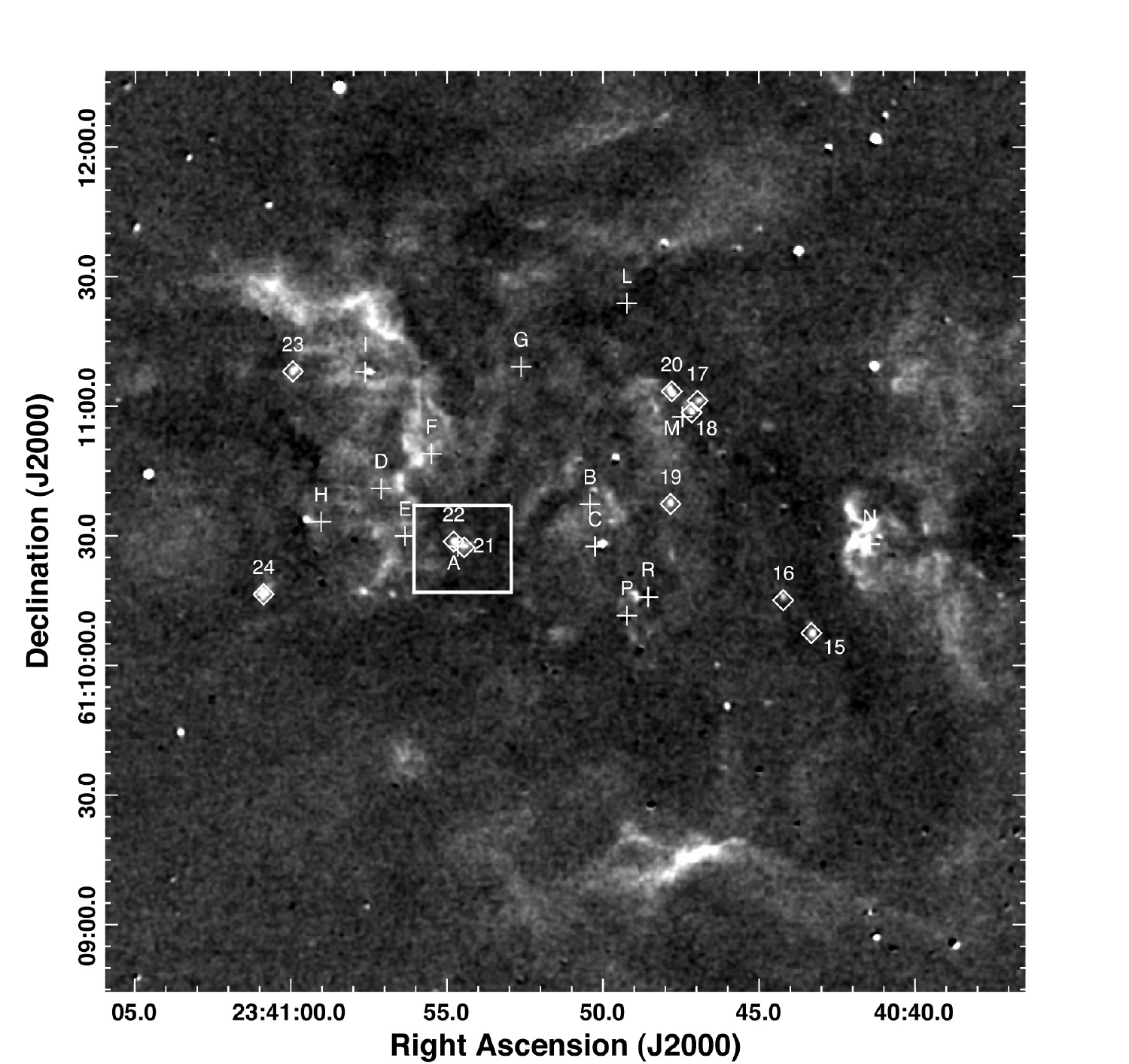}
\end{figure}
 
\clearpage
\begin{figure}
\includegraphics[angle=90,scale=0.3]{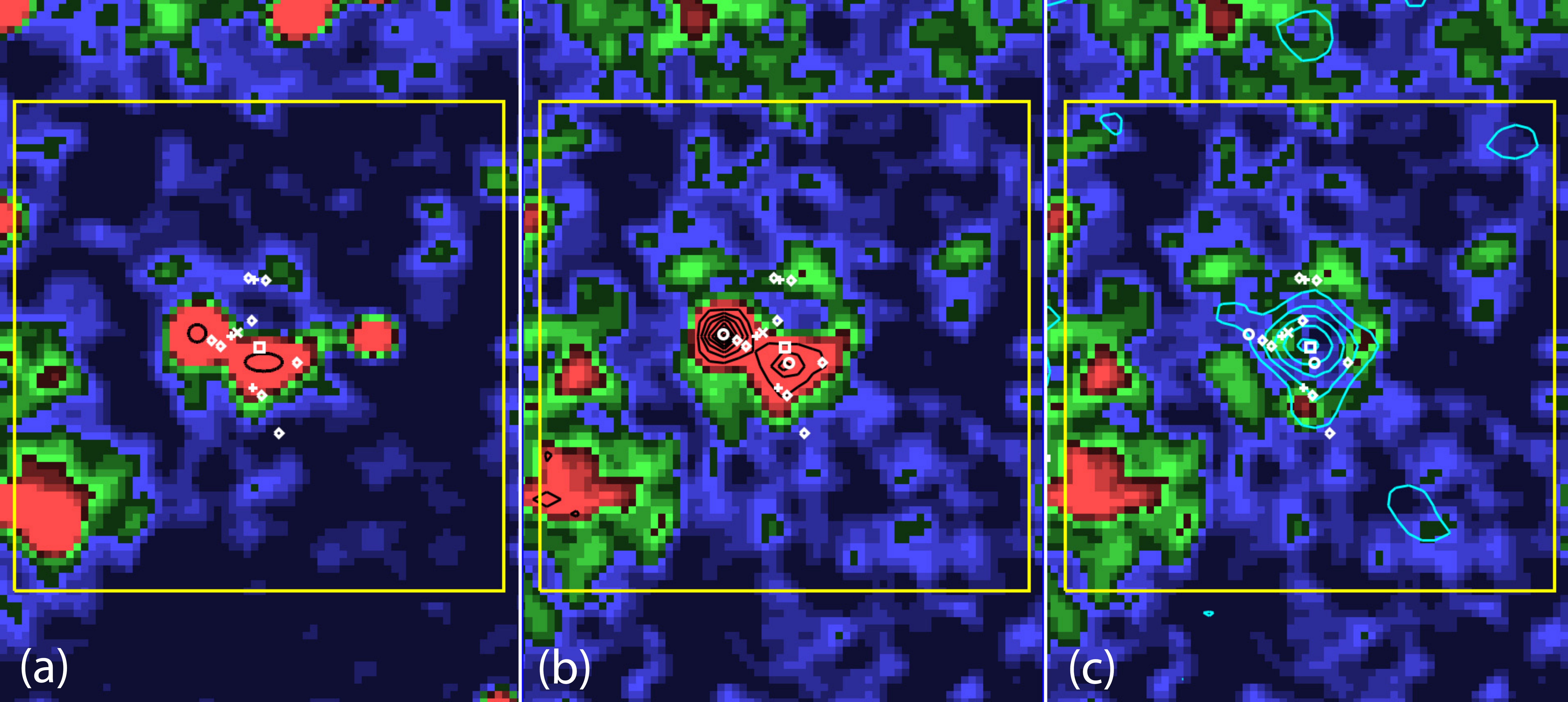}
\end{figure}

\clearpage
\begin{figure}
\includegraphics[angle=90,scale=.18]{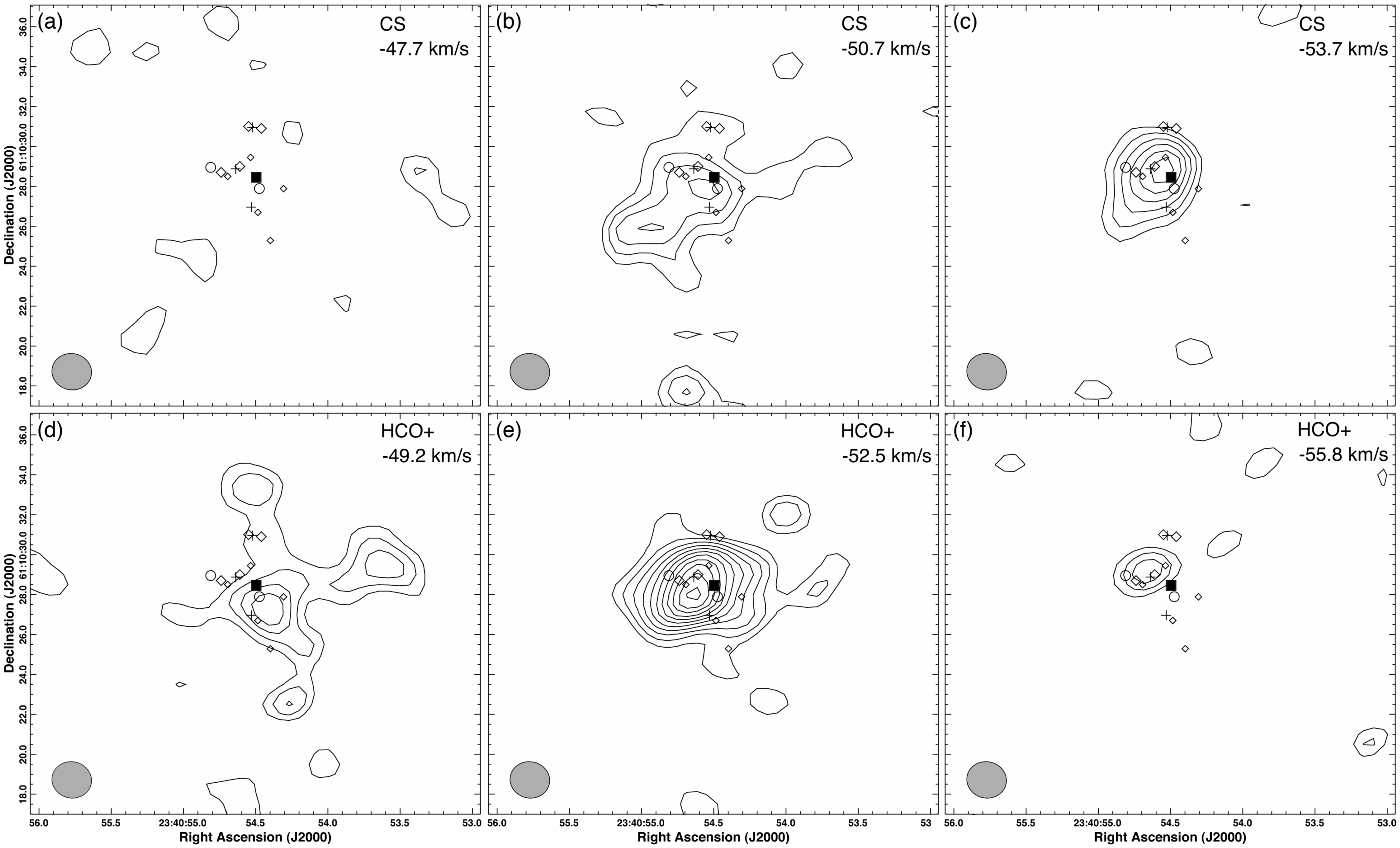}
\end{figure}

\clearpage
\begin{figure}
\includegraphics[angle=90,scale=.15]{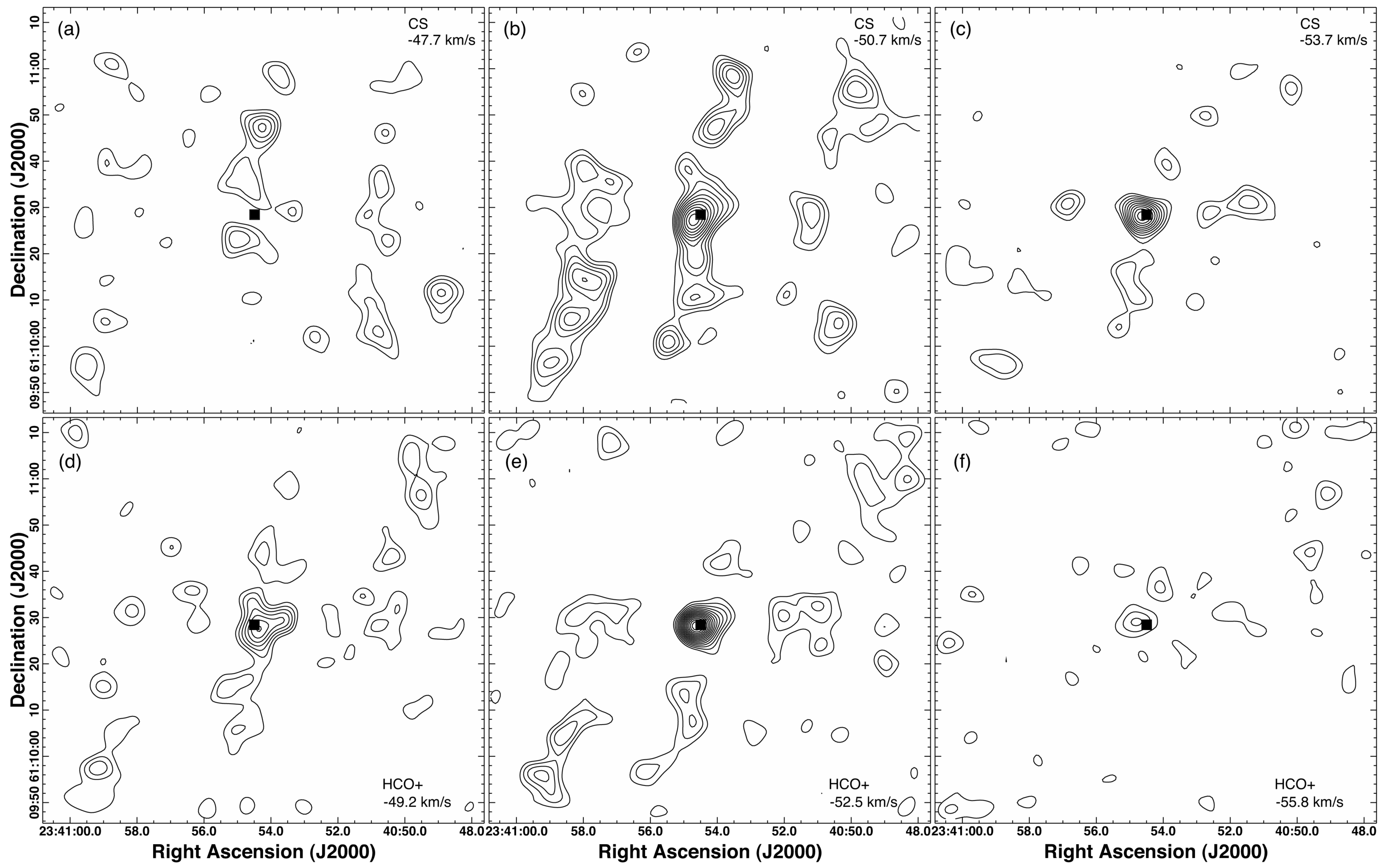}
\end{figure}

\clearpage
\begin{figure}
\includegraphics[angle=90,scale=.20]{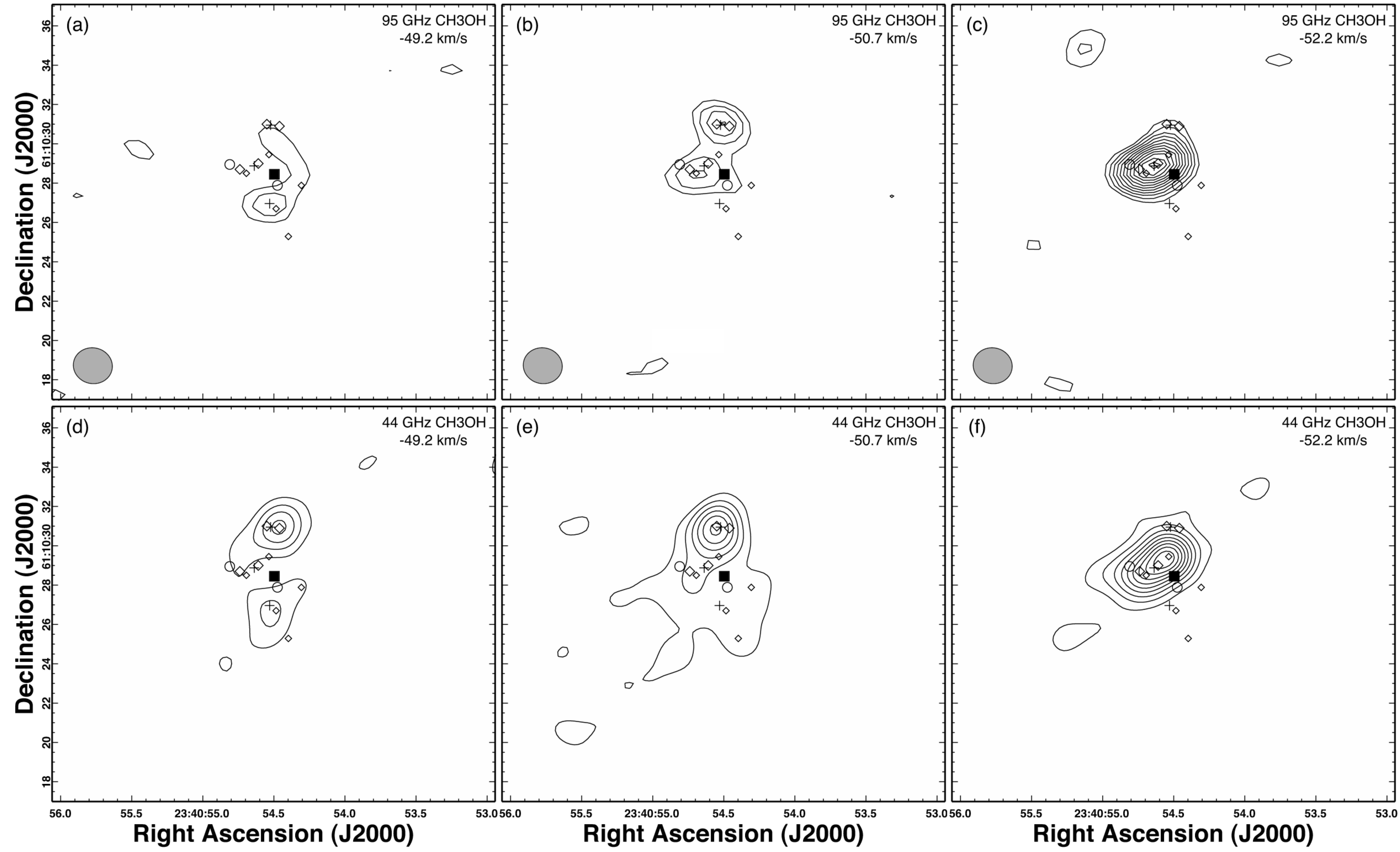}
\end{figure}

\clearpage
\begin{figure}
\plotone{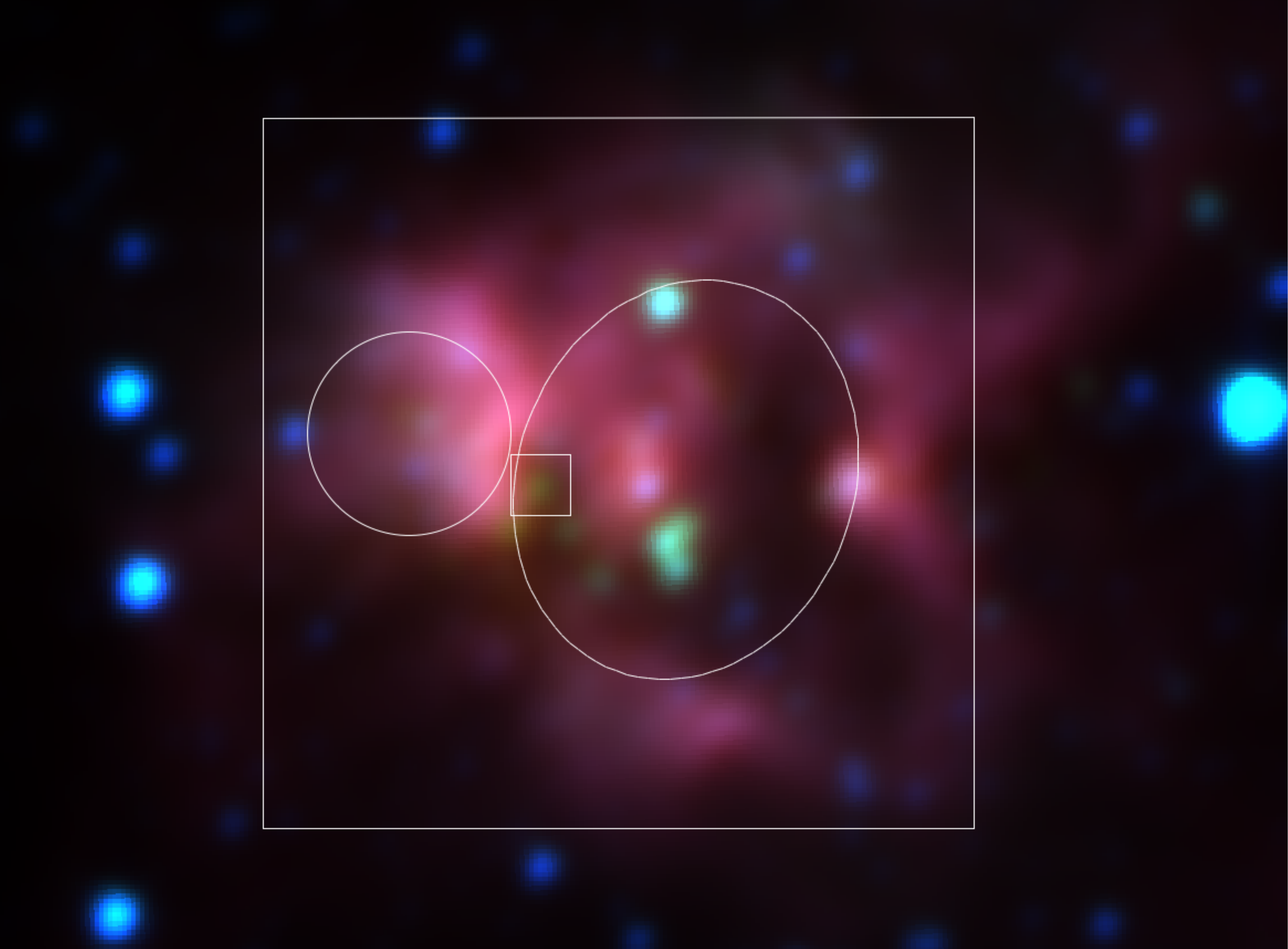}
\end{figure}

\clearpage
\begin{figure}
\includegraphics[scale=0.35]{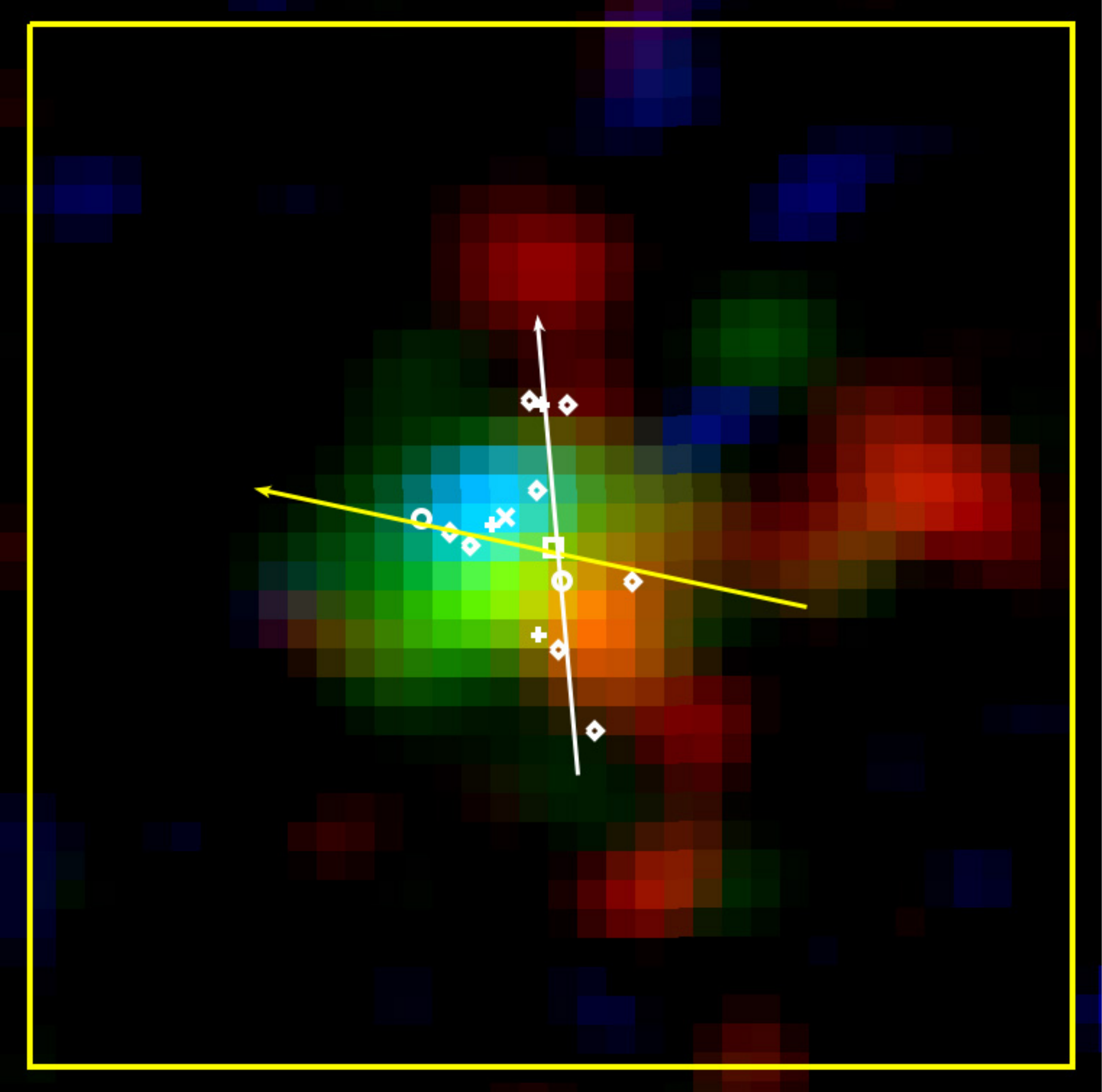}
\end{figure}

\clearpage

\begin{deluxetable}{ccccc}
\footnotesize
\tablecaption{Mol 160 MHOs and Fluxes}
\label{table1}
\tablewidth{0pt}
\tabletypesize{\scriptsize}
\tablehead{
MHO \# & $\alpha$(J2000)  & $\delta$(J2000) & 
F$_{2.12 \mu m}$ \tablenotemark{a} &  F$_{2.12 \mu m}$/F$_{2.25 \mu m}$ 
\tablenotemark{b} \\
      &  h m s     & $^{\circ}$ $^{\prime}$ $^{\prime\prime}$ & 
      10$^{-18}$ W m$^{-2}$ &   \\
      }
\startdata
2915 & 23$^h$40$^m$43.3$^s$   & 61$^{\circ}$10$^{\prime}$08$^{\prime\prime}$  &  
1.800$\pm$0.166 & $>$1.897 \\
2916 & 23$^h$40$^m$44.2$^s$ & 61$^{\circ}$10$^{\prime}$15$^{\prime\prime}$   & 
0.953$\pm$0.123 & $>$1.004  \\
2917 & 23$^h$40$^m$47.0$^s$ & 61$^{\circ}$11$^{\prime}$01$^{\prime\prime}$   &   
1.326$\pm$0.134 & $>$1.397  \\
2918 &  23$^h$40$^m$47.2$^s$  & 61$^{\circ}$10$^{\prime}$59$^{\prime\prime}$   &  
1.705$\pm$0.173  & $>$1.797 \\
2919 & 23$^h$40$^m$47.8$^s$  &  61$^{\circ}$10$^{\prime}$37$^{\prime\prime}$ & 
1.837$\pm$0.169  &  $>$1.936 \\
2920 &  23$^h$40$^m$47.8$^s$  & 61$^{\circ}$11$^{\prime}$04$^{\prime\prime}$  & 
2.631$\pm$0.242  &  $>$2.772 \\
2921 & 23$^h$40$^m$54.4$^s$  & 61$^{\circ}$10$^{\prime}$27$^{\prime\prime}$  &  
2.302$\pm$0.212  & $>$2.426  \\
2922 & 23$^h$40$^m$54.8$^s$  &  61$^{\circ}$10$^{\prime}$29$^{\prime\prime}$ &    
2.501$\pm$0.207  &  $>$2.635 \\
2923 &  23$^h$41$^m$00.0$^s$  & 61$^{\circ}$11$^{\prime}$08$^{\prime\prime}$   & 
2.406$\pm$0.221 &  $>$2.535  \\
2924 & 23$^h$41$^m$00.9$^s$ & 61$^{\circ}$10$^{\prime}$17$^{\prime\prime}$   &   
3.827$\pm$0.317  & $>$4.033 \\
\enddata
\tablenotetext{a}{Flux errors include photon counting, zero point conversion between
H$_2$ and 2MASS K$_s$, and variations in measurements of the source across different
parts of the detector.}
\tablenotetext{b}{Based on 3$\sigma$ upper limit F$_{2.25\micron} < 9.49 \times 10^{-19}$ W m$^{-2}$.}

\end{deluxetable}

\clearpage

\begin{deluxetable}{cccccccccc}
\tabletypesize{\scriptsize}
\rotate
\footnotesize
\tablecaption{Comparison of 95 \& 44-GHz CH$_3$OH Maser Properties}
\label{table 2}
\tablewidth{0pt}
\tablehead{
Source  & $\alpha$(J2000)\tablenotemark{a}   & $\delta$(J2000)\tablenotemark{a}  & 
 \# Ch\tablenotemark{b}  & 44-GHz V$_{LSR}$\tablenotemark{c}  & 
95-GHz V$_{LSR}$\tablenotemark{d}   &  $S_{44} \Delta v$\tablenotemark{e}  
&  $\Sigma S_{44} \Delta v$\tablenotemark{f} 
 &  $S_{95} \Delta v$\tablenotemark{g}    &  $S_{95}/S_{44}$\tablenotemark{h}  \\
 &   &  &  &   (km s$^{-1}$)   &  (km s$^{-1}$)   &  (mJy km s$^{-1}$) &  
 (mJy km s$^{-1}$)  &  (mJy km s$^{-1}$)  &   \\
      }
\startdata
 1  &  
23$^h$40$^m$54.46$^s$   & 61$^{\circ}$10$^{\prime}$30.92$^{\prime\prime}$ 
&  3 & -49.67   & -49.2  &  109 &  158  &  192 &  1.2  \\
 2\tablenotemark{i}  &  
23$^h$40$^m$54.55$^s$   & 61$^{\circ}$10$^{\prime}$31.02$^{\prime\prime}$ 
& 6  &  -51.00   & -50.7 & 69  &  209  &  588  & 2.8   \\
 3N  &  
23$^h$40$^m$54.54$^s$   & 61$^{\circ}$10$^{\prime}$29.44$^{\prime\prime}$ 
& 4  &  -52.00  &  -52.2 &  157  &  322 & 1341 & 4.2   \\
3E  & 23$^h$40$^m$54.69$^s$   & 61$^{\circ}$10$^{\prime}$28.49$^{\prime\prime}$  
 & 3  & -52.16  &  -52.2  & 61  & 245   & 1420 &  5.8 \\
4\tablenotemark{j}   & 23$^h$40$^m$54.74$^s$   
& 61$^{\circ}$10$^{\prime}$28.71$^{\prime\prime}$  
& 2   & -51.66  & -52.2   &  60  &  229  & 437  &  6.1  \\
4\tablenotemark{j}   & 23$^h$40$^m$54.74$^s$   
& 61$^{\circ}$10$^{\prime}$28.71$^{\prime\prime}$  
&  1  & -51.50  &  -50.7 & \nodata  &  72  &  468 & 5.4   \\
5   &  23$^h$40$^m$54.49$^s$  & 61$^{\circ}$10$^{\prime}$26.69$^{\prime\prime}$  
& 2  &   -49.01   &  -49.2 &  39  &  86  & 468  & 5.4  \\
6   &  23$^h$40$^m$54.40$^s$  & 61$^{\circ}$10$^{\prime}$25.29$^{\prime\prime}$  
& 2  &  -50.67   &  -50.7  & 30  &  51 & 126  & \nodata   \\
7   &  23$^h$40$^m$54.30$^s$  & 61$^{\circ}$10$^{\prime}$27.86$^{\prime\prime}$  
&  2  & -50.50   &  -50.7  & 22  & 55  &  \nodata  &  \nodata  \\
\enddata
\tablenotetext{a}{Position of peak intensity of the 44-GHz source.} 
\tablenotetext{b}{Number of 0.166 km s$^{-1}$ wide channels in which 44-GHz source
is present.}
\tablenotetext{c}{Central V$_{LSR}$ of the 0.166 km s$^{-1}$  wide channel in which the
44-GHz source has maximal intensity.}
\tablenotetext{d}{Central V$_{LSR}$ of the 1.538 km s$^{-1}$ wide channel 
of the 95-GHz and velocity-integrated 44-GHz source.}
\tablenotetext{e}{Per beam intensity of the brightest 0.166  km s$^{-1}$  wide channel
of the 44-GHz source.}
\tablenotetext{f}{Per beam 44-GHz intensity integrated over the 95-GHz velocity bandwidth.}
\tablenotetext{g}{Per beam  intensity of the 95-GHz source.}
\tablenotetext{h}{Ratio of 95-GHz to 44-GHz intensities (column 9 divided by column 8.)}
\tablenotetext{i}{Source 2 displays strong and systematic shifts in position across 6 channels
of the 44-GHz data.}
\tablenotetext{j}{Source 4 is listed twice because the 44-GHz channels in which it is present
overlap a boundary of two 95-GHz channels.}
\end{deluxetable}

\clearpage
\figcaption[Figure1.pdf]{Greyscale image of H$_2$ 2.12-$\mu$m emission in the Mol 160 region prior to continuum subtraction.  Large ovals indicate boundaries of the \ion{H}{2} regions (Molinari et al. 2002). Contours show velocity-integrated CS emission smoothed with an 4$^{\prime\prime}$-radius Gaussian kernel. The box delineates the region displayed in  Figures 4, 6, and 8.
}

\figcaption[Figure2.pdf]{Greyscale image of  continuum-subtracted H$_2$ 2.12-$\mu$m emission in the Mol 160 region. 
Crosses with letters indicate the positions of 24-$\mu$m point sources identified by M08. Open diamonds indicate MHOs, which are labelled using the last two digits of the MHO designations presented in Table 1.
The box delineates the region displayed in  Figures 4, 6, and 8.
}

\figcaption[Figure3.pdf]{Images of H$_2$ 2.12-$\micron$ emission in the vicinity of MHOs 2921 \& 2922 in the Mol 160 core: (a) H$_2$ 2.12-$\micron$ prior to continuum subtraction; (b) continuum-subtracted H$_2$ 2.12-$\micron$; and (c) residual H$_2$ 2.12-$\micron$ emission after subtraction of Gaussian fits to MHOs 2921 \& 2922, as described in the text. The box delineates the region displayed in  Figures 4, 6, and 8. The images have been smoothed with a 1$^{\prime\prime}$-radius Gaussian kernel to emphasize faint extended emission. The black ellipses in panel (a) show the de-convolved sizes of the MHOs. Contours of the full-resolution, non-continuum-subtracted H$_2$ image  are shown in panel (b). To show structure more clearly, contours increase in steps that are 25\% of the peak emission of MHO 2921 and 9\% of the peak emission of MHO 2922.  Contours of the 3.2-mm continuum data are shown in panel (c). Circle points indicate peaks of MHO 2921 \$ 2922 in panels (b) and (c). In all panels, the box-point shows the position of the 3.2-mm continuum peak; crosses indicate 95-GHz CH$_3$OH maser peaks; and diamonds indicate the 44-GHz CH$_3$OH maser positions listed in Table 2. The x-point indicates the KHA position for Source 3 (resolved in Table 2 into two spatial components).
}

\figcaption[Figure4.pdf]{Channel maps of CS and HCO$^+$ emission. Radial velocities are indicated in the individual panels. The noise levels of the CS and HCO$^+$ maps are, respectively, 47 and 41 mJy beam$^{-1}$. The contours begin at 2$\sigma$ and increase in 1$\sigma$ increments. The box-point shows the position of the 3.2-mm continuum peak; circles indicate the MHO peaks; crosses indicate 95-GHz CH$_3$OH maser peaks; large diamonds indicate the 44-GHz CH$_3$OH maser positions given by KHA. Small diamonds indicate the additional 44-GHz maser positions listed in Table 2.
}

\figcaption[Figure5.pdf]{Channel maps of CS and HCO$^+$ emission, smoothed with a 4$^{\prime\prime}$-radius Gaussian kernel to emphasize extended emission. The box-point shows the position of the 3.2-mm continuum peak.
}

\figcaption[Figure6.pdf]{Channel maps of 95-GHz CH$_3$OH emission and of 44-GHz CH$_3$OH emission integrated over the same velocity range as the 95-GHz channels and smoothed to the same angular resolution. For the 95-GHz channels, contours start at 2$\sigma$ and increment by 1$\sigma$. For the 44-GHz channels, contours start at 20 mJy/beam and increment by 20 mJy/beam ($\sim$ 2$\sigma$). Symbols are the same as for Figure 4.
}

\figcaption[Figure7.pdf]{Three-color WISE image of the Mol 160 region. Colors are 3.4 $\micron$ (blue), 4.6 $\micron$ (green), and 12 $\micron$ (red). Large ovals indicate boundaries of the \ion{H}{2} regions (Molinari et al. 2002). The smaller box indicates the region displayed in Figures 4, 6, and 8. The larger box indicates the region displayed in Figures 1 \& 2.
}

\figcaption[Figure8.pdf]{Color image made by combining the HCO$^+$ channels from  Figure 4. Colors are -49.2 km s$^{-1}$ (red), -52.5 km s$^{-1}$ (green), and -55.8 km s$^{-1}$ (blue). Symbols are the same as for Figure 3, panels (b) and (c). The arrows are drawn through the continuum peak along possible preferred axes of maser activity.
}

\end{document}